\documentclass[preprint,12pt]{elsarticle} 
\usepackage{amsmath,amsfonts,amssymb}
\usepackage{algorithmic}
\usepackage{algorithm}
\usepackage{array}
\usepackage[caption=false,font=normalsize,labelfont=sf,textfont=sf]{subfig}
\usepackage{textcomp}
\usepackage{stfloats}
\usepackage{url}
\usepackage{verbatim}
\usepackage{graphicx}
\usepackage{tikz}
\usetikzlibrary{positioning}
\usepackage{balance}

\newcommand{\substarget}{\mathbf{x}}
\newcommand{\subssensor}{\mathbf{s}}
\newcommand{\pos}{\mathbf{p}}
\newcommand{\velo}{\mathbf{v}}

\newcommand{\obs}{\mathbf{y}}
\newcommand{\meanob}{\boldsymbol{\mu}}
\newcommand{\covob}{\boldsymbol{\Sigma}}
\newcommand{\transtarget}{\mathbf{F}}
\newcommand{\transadv}{\mathbf{G}}

\newcommand{\control}{\mathbf{u}}

\newcommand{\seqctr}{\mathbf{U}}
\newcommand{\setctr}{\mathcal{U}}

\newcommand{\adversarial}{\mathbf{a}}
\newcommand{\covadv}{\mathbf{Q}}
\newcommand{\estimator}{\widehat{\mathbf{x}}}
\newcommand{\predictor}{\check{\mathbf{x}}}
\newcommand{\step}{t}

\newcommand{\costfunc}{\mathcal{J}}
\newcommand{\coverror}{\mathbf{C}}
\newcommand{\covprederror}{\check{\coverror}}

\newcommand{\weight}{\mathbf{W}}
\newcommand{\fim}{\mathbf{J}}

\newcommand{\identitymat}{\mathbf{I}}
\newcommand{\zeromat}{\mathbf{0}}

\newcommand{\T}{\intercal}

\DeclareMathOperator*{\argmin}{arg\,min}
\newcommand{\z}{\mathbf{z}}
\newcommand{\boundctr}{u_{\text{max}}}
\newcommand{\boundadv}{a_{\text{max}}}

\newcommand{\trajectory}{\mathbf{z}}

\begin{document}
\begin{frontmatter}
\title{Target Tracking using Robust Sensor Motion Control}
\author{Jingwei Hu \quad  Dave Zachariah \quad Petre Stoica}


\begin{abstract}
We consider the problem of tracking moving targets using mobile wireless sensors (of possibly different types). This is a joint estimation and control problem in which a tracking system must take into account both target and sensor dynamics. We make minimal assumptions about the target dynamics, namely only that their accelerations are bounded. We develop a control law that determines the sensor motion control signals so as to maximize target resolvability as the target dynamics evolve. The method is given a tractable formulation that is amenable to an efficient search method and is evaluated in a series of experiments involving both round-trip time based ranging and Doppler frequency shift measurements.
\end{abstract}
\end{frontmatter}
\section{Introduction}

Tracking the positions of moving objects (aka. targets) using wireless signals is a classical estimation problem \cite{barshalom,kailath2000linear}, in which the ability to track crucially depends on the spatial configuration of targets and sensors. The targets can be unmanned aerial vehicles, aircraft or land vehicles. The development of wireless technologies has enabled the flexible deployment of sensors and,  consequently, the potential of finding favourable sensor configurations in a given application. 

In the case of \emph{static targets}, one important line of work is that of optimal sensor placement, which uses the Fisher Information Matrix (FIM) and associated metrics to quantify the ability to resolve a target  \cite{yangsensorplacement,NeeringPassiveLocalization,JUHLIN2023108679,xu2022hybrid}. Sensors that are mobile, or in other ways dynamically reconfigurable, can be used to improve target localization sequentially. This is illustrated in \cite{Oshman-fim-cooperative-tracking}, which considers the problem of designing a control law for a single bearings-only sensor. The proposed control law is based on maximizing the determinant of the (estimated) FIM using a first-order method. An alternative case with multiple ranging sensors is considered in \cite{MARTINEZ2006661}, where a distributed algorithm is developed for steering the sensors along the boundary of a convex set containing the target.

In the case of \emph{dynamic targets}, the optimal sensor configurations change dynamically. In \cite{cooperative-tracking-control}, a fuzzy-logic based method is developed for distributed sensor control using point estimates of a single target. The estimates are obtained using distributed Kalman filters under the assumption of a linear measurement model.  Tracking using mobile sensors requires an appropriately designed control law, since their motion towards a currently favourable configuration may become unfavourable at the time they reach it. This includes taking into account estimation errors, as exemplified in \cite{active-target-tracking} which considers the problem of tracking using aerial vehicles equipped with position measuring sensors. In the cited paper, sensor control is based on reducing the estimated error covariance matrix obtained from a Kalman filter using a first-order method. A robust approach that takes into account the least favourable target trajectories as well as measurement noise realizations is developed in \cite{tree-minimax-tracking-control}. The sensor control uses a receding horizon formulation to minimize the trace of the Kalman filter error covariance matrix at the end of the time horizon. The control law is implemented by gridding the target action and noise realization spaces, and then performing a minimax tree search. While increasing robustness, the computational complexity of the minimax tree grows prohibitively with the time horizon, even when utilizing gridding and pruning techniques. 

In this work, we develop a  tracking framework to determine a robust control law for mobile sensors that
\begin{itemize}
    \item is applicable to various (and possibly mixed) sensor types,
    \item employs minimal assumptions about the targets,
    \item aims at improving target resolvability, and
    \item can be implemented efficiently using first-order methods.
\end{itemize}
We formulate the sensor control law that aims to minimize an expected Cramér-Rao bound (CRB) using a computationally efficient first-order method.

\begin{figure*}[!htb]
    \centering
    \subfloat[]{\includegraphics[width=0.45\linewidth]{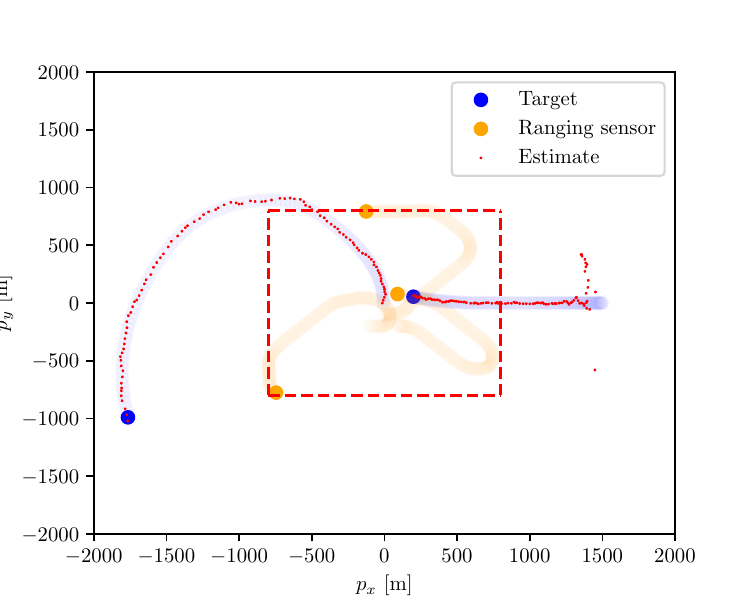}\label{fig:two-target-ranging-sensor}}
    \subfloat[]{\includegraphics[width=0.45\linewidth]{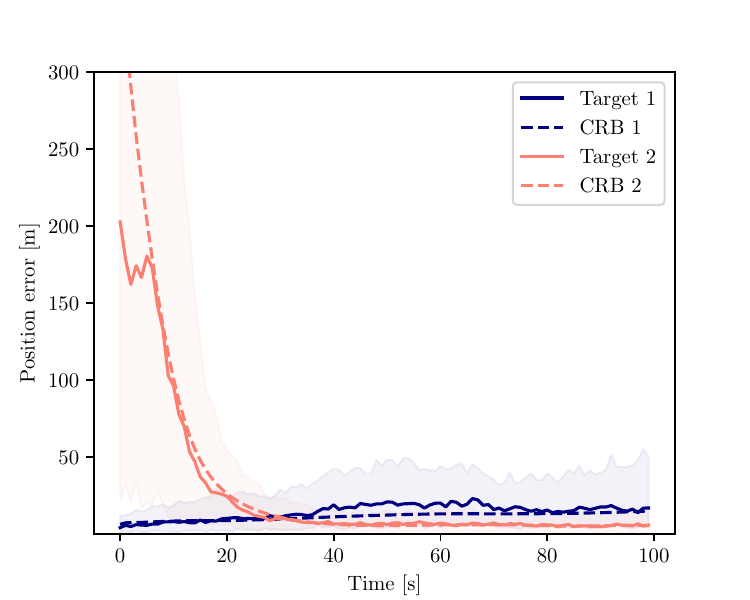}\label{fig:two-target-range-sensor-err}}
    \caption{Proposed tracking system using sensor motion control. $N=2$ targets are tracked in two-dimensional space ($d=2$) using  $M=3$ sensors that measure the ranges to the targets. Target and sensor accelerations are assumed to be bounded by
    $\boundadv= 5 \text{ m/$\text{s}^2$}$ and $\boundctr = 2 \text{ m/$\text{s}^2$}$, respectively. The sensor errors are specified in Section~\ref{sec:measurementmodels}. (a) The targets move in an unknown, coordinated manner. The sensors are initially positioned at the center of the perimeter and can move freely within this boundary. One target moves from the center of the perimeter toward the outside, while the other moves inward, entering the perimeter. (b) The resulting tracking position errors over time. For reference we have included the Cramér-Rao (CRB) bounds which show that the errors are driven down to their respective `noise floors' determined by the sensor configuration. Details about the experiments are specified in Section~\ref{sec:experiments}.}
\label{fig:openingexperiment}
\end{figure*}

Figure~\ref{fig:two-target-ranging-sensor} demonstrates the proposed framework for tracking two targets using three ranging sensors that are confined to a rectangular perimeter. The tracking system uses a maximum likelihood estimator and only assumes an upper bound  $\boundadv$ on the accelerations of the targets, which move in an unknown coordinated manner.  The sensors are initialized at the center of the perimeter resulting in poor localization capabilities.
Figure~\ref{fig:two-target-range-sensor-err} shows the position errors over time. We can see that after overcoming the challenging sensor initialization, the system drives the errors down towards their corresponding `noise floors' as indicated by the corresponding CRB.

Next, we turn to setting up the formal problem that this paper addresses. We then turn to developing our proposed joint estimation and control method. Finally, a statistical evaluation is provided for a few different scenarios with aerial vehicles using mixed sensor types.

\section{PROBLEM FORMULATION}

We consider the problem of tracking $N$ targets using $M$ mobile sensors. Consider a target $n$ observed by a sensor $m$. Their positions and velocities in $d$-dimensional space ($d=2$ or $3$) are \emph{states} denoted
\begin{equation}
\substarget^n = \begin{bmatrix} \pos^n \\ \velo^n\end{bmatrix} \quad \text{and} \quad  \subssensor^m = \begin{bmatrix} \widetilde{\pos}^m \\ \widetilde{\velo}^m\end{bmatrix},
\label{eq:states_def}
\end{equation}
respectively.

At time $t$, sensor $m$ provides a measurement of target $n$, which we denote $y^{m,n}_{t}$. The sensors may be of different types, providing, for instance, round-trip time or frequency shift measurements, which we will consider below. In the case of ranging measurements, the mean round-trip time from sensor $m$ to target $n$, and back, is
\begin{equation}
\mathbb{E}[y^{m,n}_{t}] = \frac{2}{c} \| [\identitymat_d \; \zeromat] (\substarget^n_t - \subssensor^m_t)  \|_2,\label{eq:ranging}
\end{equation}
where $c$ is the speed of light.
In the case of Doppler frequency shift measurements, the mean observation is
\begin{equation}
\mathbb{E}[y^{m,n}_{t}]  = -\frac{f_c}{c}\frac{\sum_{i=1}^d (x^n_i-s^m_i)(x^n_{d+i}-s^m_{d+i})}{\| [\identitymat_d \; \zeromat] (\substarget^n - \subssensor^m)  \|_2},\label{eq:doppler}
\end{equation}
where $f_c$ is the carrier frequency \cite{doppler} and the index $i$ denotes the $i$th element of the target and sensor.  

Both target $n$ and sensor $m$ can be in motion, which we describe using a general discrete-time model:
\begin{equation}
\substarget^n_{t}= \underbrace{\begin{bmatrix}
\identitymat & d\step \identitymat \\
\zeromat & \identitymat
\end{bmatrix}}_{\transtarget} \substarget^n_{\step-1} + \underbrace{\begin{bmatrix} \frac{1}{2} d\step^2\identitymat  \\ d\step \identitymat \end{bmatrix}}_{\transadv}  \adversarial^n_{\step},
\label{eq:targetdynamics}
\end{equation}
where $dt$ denotes the sampling period, $\adversarial^n$ is the acceleration vector of the target
and
\begin{equation}
\subssensor^m_{\step} = \underbrace{\begin{bmatrix}
\identitymat & d\step \identitymat \\
\zeromat & \identitymat
\end{bmatrix}}_{\widetilde{\transtarget}} \subssensor^m_{\step-1} + \underbrace{\begin{bmatrix} \frac{1}{2}d\step^2\identitymat \\ d\step \identitymat \end{bmatrix}}_{\widetilde{\transadv}} \control^m_{\step},
\label{eq:sensordynamics}
\end{equation}
where $\control^m$ is the acceleration vector of the mobile sensor. The trajectory of the target is unknown. For the sake of robustness, we will make \emph{minimal} assumptions and  consider only physical constraints on its unknown acceleration:
\begin{equation}
\| \adversarial^n_t \|_2 \leq \boundadv.
\label{eq:targetconstraint}
\end{equation}
The mobile sensor acceleration is a
\emph{control input} that we can design subject to certain constraints: Firstly, it is bounded 
\begin{equation}
\| \control^m_t \|_2 \leq \boundctr.
\label{eq:sensorconstraint}
\end{equation}
and, secondly, the sensor must often stay within a certain perimeter or respect certain velocity limits described by the set of box constraints
\begin{equation}
 \{ \subssensor^m_t : \subssensor_{\min} \leq \subssensor^m_t \leq \subssensor_{\max}, \: \forall t \}.\label{eq:boxconstraint}
\end{equation}

The problem we face is to coordinate the motion of $M$ sensors so as to track target all $N$ accurately. More specifically, given measurement obtained from all sensors at time $t$, our aim is to design the subsequent control input $\control^m_{\step+1}$
so that a target state estimator
$\estimator^n_{\step+1}$ will perform well for all $n$. 

\section{Method}
 For notational convenience, we introduce the sensor state \emph{configuration}:
\begin{equation}
\subssensor_{t} = \begin{bmatrix}
\subssensor^1_{t} \\ \vdots \\ \subssensor^M_{t}
\end{bmatrix} 
\label{eq:controlinputs}
\end{equation}

Our tracking method is based on the maximum likelihood estimator (MLE). Let $\obs^n_{t}$ denote the vector of all measurements of target $n$ at time $t$ and assume it follows a Gaussian data model $\obs^n_{t} \sim \mathcal{N}( \meanob(\substarget^n,\subssensor_{t}), \covob(\substarget^n,\subssensor_{t}) )$, where the mean and covariance depend both on the known sensor configuration and the unknown target state. Suppose we have an unbiased prediction $\predictor^n_{t}$ with an error covariance matrix $\covprederror_{n,t}$. Using $\obs^n_{t}$ and $\predictor^n_{t}$ as two sources of data, we have the maximum likelihood estimator of the target state:
\begin{equation}
\begin{split}
\widehat{\substarget}^n_\step &= \argmin_{\substarget^n} \: \| \obs^n_{t} - \meanob(\substarget^n,\subssensor_{t})\|^2_{\covob^{-1}(\substarget^n,\subssensor_{t})} + \ln |\covob(\substarget^n,\subssensor_{t})| \\
&\quad +  \| \substarget^n - \predictor^n_t\|^2_{\covprederror^{-1}_{n,\step}}, \quad n=1, \dots, N,
\end{split}
\label{eq:mle}
\end{equation}
where we have used the least favorable distribution  $\predictor^n_{t} \sim \mathcal{N}( \substarget^n_t, \covprederror_{n,t} )$ \cite{stoica2011gaussian,park2013gaussian}. The error covariance matrix of the MLE $\widehat{\substarget}^n_\step$ is approximated by
\begin{equation}
  \coverror_{n, t} =
\big( \fim(\substarget^n_{t}, \subssensor_{t}) + \covprederror^{-1}_{n,t} \big)^{-1},
\label{eq:mle_coverror}
\end{equation}
when applying a plug-in estimate of the target state \cite{van1968detection, kay1993, stoica2005spectral}. Here $\fim$ is the Fisher information matrix given by
\begin{equation*}
\begin{split}
[\fim(\substarget^n,\subssensor)]_{ij} &= \frac{\partial{\meanob(\substarget^n,\subssensor)}}{\partial{x}^n_i}^\top \covob^{-1}(\substarget^n,\subssensor) \frac{\partial{\meanob(\substarget^n,\subssensor)}}{\partial{x}^n_j}\\ &+ \frac{1}{2}\text{tr}\left\{\covob^{-1}(\substarget^n,\subssensor)\frac{\partial{\covob(\substarget^n,\subssensor)}}{\partial{x}^n_i}\covob^{-1}(\substarget^n,\subssensor)\frac{\partial\covob(\substarget^n,\subssensor)}{\partial{x}^n_j}\right\}, 
\end{split}
\end{equation*} 
for $i,j =1, \dots, 2d $ (and dropping time index $t+k$ for notational brevity).

We consider $\predictor^n_{t}$ to be a $k$-step prediction. This can be formed using the dynamic model \eqref{eq:targetdynamics}:
Suppose the unknown $\adversarial^n_t$ in \eqref{eq:targetdynamics} is modeled as a zero-mean random variable with a covariance matrix $\covadv(\boundadv)$. Using a uniform distribution for $\adversarial^n_t$ over $[-\boundadv, \boundadv]^d$ yields conservative relaxation of the bound \eqref{eq:targetconstraint} and a covariance matrix
 $$\covadv(\boundadv) = \frac{\boundadv^2}{3} \identitymat_{d}.$$ Then starting at a prior estimate at time $t-k$, we have the following $k$-step predictor, $\predictor^n_{t} = \transtarget^k \estimator^n_{t-k}$  with a resulting error covariance matrix (using \eqref{eq:sensordynamics}, see \cite{kailath2000linear})
\begin{equation}
    \covprederror_{n,\step} =  \transtarget^{k} \coverror_{n,t-k} (\transtarget^k)^\T + \sum^{k}_{i=1} \transtarget^{i-1}  \transadv \mathbf{Q} \transadv^\T (\transtarget^{i-1})^\T ,
\label{eq:predictor_fim}
\end{equation}
where $\coverror_{n,t-k}$ is prior error covariance approximated by \eqref{eq:mle_coverror}.

\begin{figure}
    \centering
    \begin{tikzpicture}
\node [draw,
    minimum width=1cm, 
    minimum height=1cm, 
]  (sensor) {Sensors};
\node [draw,
    minimum width=1.3cm, 
    minimum height=1cm, 
    right=1cm of sensor
]  (estimator) {Estimator};

\draw[-latex] (sensor.east) -- (estimator.west) 
    node[midway,above](y){$\mathbf{y}$};

\node [draw,
    minimum width=1.3cm, 
    minimum height=1cm, 
    below=1.48 cm of y
]  (controller) {Controller};

\draw[-latex](estimator.east) -- ++ (1.5,0) 
    node[midway,above](output){$\hat{\mathbf{x}},\mathbf{C}$};

\draw[-latex] (output) |- (controller.east)
    node[midway,above]{};

\draw[-latex] (controller.west)-| ++(-2, 2) -- (sensor.west)
    node[midway,above]{$\mathbf{u}$};
\end{tikzpicture}
    \caption{Tracking method with sensor motion control. The sensors provide a snapshot of measurements $\obs$. This is fed to an estimator which provides a point estimate $\estimator$ of the target states and an error covariance matrix, $\coverror$. Based on these two quantities,s the controller determines the sensor inputs $\control$ for the next time step.} 
    \label{fig:block_diagram}
\end{figure}
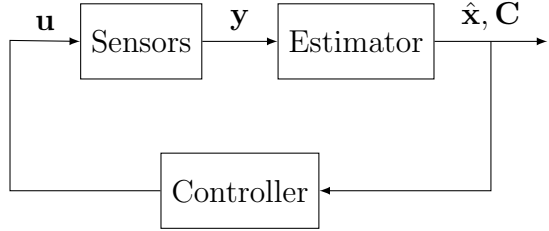

An overview of the tracking framework proposed below is illustrated in Figure \ref{fig:block_diagram}.

\subsection{Sensor state  configuration and control objective}

To plan the sensor trajectories at time $t$, we are interested in quantifying how the resulting sensor state configuration $\subssensor_{t+k}$ at time $t+k$ will affect the estimation error, defined as
\begin{equation}
\mathbb{E}\Big[ \| \substarget^n_{t+k} - \estimator^n_{t+k} \|^2_{\weight}\Big]
\label{eq:weightedmse}
\end{equation}
where $\weight \succeq \zeromat$ is a weight matrix of choice, e.g.,
\begin{equation}
\weight \propto \begin{bmatrix}
\identitymat_d & \zeromat \\
\zeromat & dt \identitymat_d
\end{bmatrix},\label{eq:fim_weight}
\end{equation}
which coverts the velocity errors at any time step into equivalent position errors. 


Using measurements $\obs^n_{t+k}$ and the $k$-step prediction $\predictor^n_{t+k}$ as data, the Cram\'{e}r-Rao inequality  \cite{van1968detection, kay1993, stoica2005spectral} yields an 
(asymptotically) achievable lower bound on the estimation error \eqref{eq:weightedmse}. Specifically, we consider the expected CRB, by treating $\substarget^n_{t+k}$ as a random variable:
\begin{equation}
\begin{split}
 \mathbb{E}\Big[ \| \substarget^n_{t+k} - \estimator^n_{t+k} \|^2_{\weight}\Big] &\geq \mathbb{E} \Big[ g( \substarget^n_{t+k},\subssensor_{t+k} ) \Big],
 \end{split}
\label{eq:crb}
\end{equation}
where
\begin{equation}
g( \substarget^n_{t+k},\subssensor_{t+k} ) \equiv \text{tr}\left\{ \weight \big( \fim(\substarget^n_{t+k}, \subssensor_{t+k}) + \covprederror^{-1}_{n,t+k} \big)^{-1} \right\},
\end{equation}
the prediction error covariance $\covprederror_{n,t+k}$ is given by \eqref{eq:predictor_fim}
and  Thus $\mathbb{E} [ g( \substarget^n_{t+k},\subssensor_{t+k} )]$ in \eqref{eq:crb} provides a $k$-step ahead error bound, as a function of the unknown target state and the sensor state configuration. It is this quantity we seek to reduce by controlling the sensor motion.

We now consider the \emph{feasible} $k$-step sensor trajectories in order to minimize the expected CRB \eqref{eq:crb}. A joint sequence of  control inputs 
\begin{equation*}
\seqctr_\step = 
\begin{bmatrix}
\control^1_{\step+1} & \cdots & \control^1_{\step+k} \\
\vdots & \ddots & \vdots  \\
\control^M_{\step+1} & \cdots & \control^M_{\step+k}
\end{bmatrix}
\end{equation*}
determines the sensor state configuration $\subssensor_{t+k}$ \eqref{eq:controlinputs} via  the dynamic model \eqref{eq:sensordynamics}. That is, for each sensor $m$,
\begin{equation}
\begin{split}
\subssensor^m_{t+k} &\equiv \widetilde{\transtarget}^{k} \subssensor^m_t + \sum^{k}_{i=1} 
 \widetilde{\transtarget}^{i-1}  \widetilde{\transadv} \control^m_{t+i}.
\end{split}
\label{eq:sensortrajectory}
\end{equation}
Similarly, the target trajectory starting at $\substarget^n_{t}$ is given by
\begin{equation}
\begin{split}
\substarget^n_{t+k} \equiv \transtarget^{k} \substarget^n_t + \sum^{k}_{i=1} 
\transtarget^{i-1} \transadv \adversarial^n_{t+i} ,
\end{split}
\label{eq:targettrajectory}
\end{equation}
where the acceleration must satisfy \eqref{eq:targetconstraint}. Note that the target trajectory is treated as a random variable.

The arguments of $g( \substarget^n_{t+k},\subssensor_{t+k} )$ in \eqref{eq:crb} can now be substituted by the target and sensor trajectory variables in \eqref{eq:sensortrajectory} and \eqref{eq:targettrajectory}:
\begin{equation}
g(\substarget^n_{t}, \adversarial^n_{t+1}, \dots, \adversarial^n_{t+k}, \seqctr_{t} ) \equiv g(\substarget^n_{t+k}, \subssensor_{t+k}),
\label{eq:crb_alt}
\end{equation}
where the target acceleration vectors must satisfy \eqref{eq:targetconstraint} while the control inputs must belong to the set of bounded accelerations
\begin{equation}
\setctr = \Big\{ \seqctr_t : \| \control^m_{t+k} \|_2 \leq \boundctr, \: \forall k,m \Big\}
\label{eq:ctrconstraint_alt}
\end{equation}
as well as
\begin{equation}
\setctr_c = \Big\{ \seqctr_t :  \subssensor_{\min} \leq \widetilde{\transtarget}^{k} \subssensor^m_t + \sum^{k}_{i=1} 
 \widetilde{\transtarget}^{i-1}  \widetilde{\transadv} \control^m_{t+i} \leq \subssensor_{\max}, \: \forall  k,m \Big\}
\label{eq:boxconstraint_alt}
\end{equation}
for the box constraints \eqref{eq:sensorconstraint}. 

We take the expected CRB \eqref{eq:crb}, together with \eqref{eq:crb_alt}, as our control objective. Specifically, we use the lower bound on the sum of estimation errors for the $N$ targets:
\begin{equation}
\begin{split}
\seqctr^*_{\step} &= \argmin_{\seqctr_\step \in \setctr \cap \setctr_c}  \costfunc(\seqctr_{\step}),
\end{split}
\label{eq:controlproblem}
\end{equation}
where
\begin{equation}
\begin{split}
\costfunc( \seqctr_\step) &= \sum^N_{n=1} \mathbb{E}[ g(\substarget^n_{t+k}, \adversarial^n_{t+1}, \dots, \adversarial^n_{t+k}, \seqctr_\step )] 
\end{split}
\label{eq:criterion}
\end{equation}
The control inputs $\control^m_{\step+1}$ for each mobile sensor at time $t$ would then be obtained by solving \eqref{eq:controlproblem}. Because the criterion takes into account possible target trajectories, the resulting control method is robust against target uncertainties. Unfortunately, as it stands, \eqref{eq:controlproblem} is an intractable problem.

\subsection{Approximation and relaxation of optimization problem}

To render \eqref{eq:controlproblem} tractable, we first turn to approximating the expectation in \eqref{eq:criterion} and then turn to relaxing the constraints in \eqref{eq:ctrconstraint_alt}.

For notational simplicity, we concatenate the sequence
$(\substarget^n_{t}, \adversarial^n_{t+1}, \dots, \adversarial^n_{t+k})$ into a single vector for dimension $d_z = (2+k)d$, denoted $\trajectory^n_{t+k}$, that determines the trajectory of target $n$. We now approximate the expectation over the target trajectory in \eqref{eq:criterion} using deterministic sigma-point sampling method \cite{wan2000unscented,candy2016bayesian}. That is, we average over $2d_z+1$ samples $\trajectory^n_{t+k,(\ell)}$ of each target trajectory:
\begin{equation}
\begin{split}
\costfunc(\seqctr_{\step}) \simeq
 \sum^N_{n=1} \sum^{2d_z+1}_{\ell=0} g(\trajectory^n_{t+k,(\ell)}, \seqctr_\step ),
\end{split}
\label{eq:criterion_approx}
\end{equation} 
where the $\ell$th sample for target $n$ is given by
\begin{equation}\label{eq:determ-sampling}
    \trajectory_{\step+k,(\ell)}^n = \begin{cases}\Bar{\trajectory}_{\step+k}^n\text{, for }\ell=0\\
   \Bar{\trajectory}_{\step+k}^n + \eta^{1/2}[\mathbf{P}_{\step+k}^{1/2}]_\ell\text{, for }\ell=1,\dots,d_z\\
   \Bar{\trajectory}_{\step+k}^n - \eta^{1/2}[\mathbf{P}_{\step+k}^{1/2}]_{\ell-d_z}\text{, for }\ell=d_z+1,\dots,2d_z+1
    \end{cases} 
\end{equation}
and
\begin{equation}
    \Bar{\trajectory}_{\step+k}^n = \begin{bmatrix}\estimator^n_\step \\ \mathbf{0} \end{bmatrix} \quad 
    \mathbf{P}^n_{\step+k} = \begin{bmatrix}
    \coverror^n_\step&\dots&\mathbf{0}\\
    \vdots&\ddots&\vdots\\
    \mathbf{0}&\dots&\covadv
\end{bmatrix}.
\end{equation}
The scaling factor $\eta$ is set according to the principles in \cite{candy2016bayesian}.

To further speed up the solution of \eqref{eq:controlproblem}, we relax the quadratic constraints $\setctr$ by replacing them with box constraints. Utilizing the fact that $ \| \control \|_2 \leq \sqrt{\text{dim}(\control)}\| \control\|_\infty$ for any vector $\z$, we can relax the constraint in \eqref{eq:sensorconstraint} into
\begin{equation*}
 \| \control^m_t \|_{\infty} \leq \frac{\boundctr}{\sqrt{d}}.
 \label{eq:constraints_relax}
\end{equation*}
We denote the corresponding set of constraints as $\setctr_\infty$.

In sum, we replace the criterion in \eqref{eq:controlproblem} by the sum \eqref{eq:criterion_approx} and the quadratic constraint $\setctr$ by  $\setctr_\infty$ so that: 
\begin{equation}
\seqctr^*_{\step} = \argmin_{\seqctr_\step \in \setctr_{\infty} \cap \setctr_c}   \sum^N_{n=1} \sum^{2d_z+1}_{\ell=0} g(\trajectory^n_{t+k,(\ell)}, \seqctr_\step ),
\label{eq:controlproblem_relax}
\end{equation}
 Since all constraints become linear, we apply a computationally efficient interior-point solver to solve \eqref{eq:controlproblem_relax}. The resulting tracking system, summarized in  Figure~\ref{fig:block_diagram}, is described in Algorithm \ref{alg1} below.

\begin{algorithm}[H]
\begin{algorithmic}
\STATE {joint\_estimate\_control}$(\obs_{\step}):$
\STATE \hspace{0.5cm}$(\estimator_{\step},\coverror_{\step}) \gets \text{estimator}(\obs_{\step})$
\STATE \hspace{0.5cm}$ \{\control^m_{\step+1}\} \gets \text{robust\_control}(\estimator_{\step},\coverror_{\step})$
\STATE \hspace{0.5cm}\textbf{return}  $ \estimator_{\step},\coverror_{\step}, \{ \control^m_{\step+1} \}$
\STATE 

\STATE {robust\_control}$(\estimator_{\step},\coverror_{\step}):$
\STATE \hspace{0.5cm} Sample trajectories from $\estimator_\step, \coverror_{\step}$ using \eqref{eq:determ-sampling}
\STATE \hspace{0.5cm} Obtain solution $\seqctr^*_{\step+1}$ to \eqref{eq:controlproblem_relax} using interior-point solver.
\STATE \hspace{0.5cm}\textbf{return}  $\{\control^{m,*}_{\step+1} \}$ 
\end{algorithmic}
\caption{Tracking with sensor motion control}
\label{alg1}
\end{algorithm}

\section{NUMERICAL EXPERIMENTS}
\label{sec:experiments}

We present a series of numerical experiments that demonstrate the ability of the proposed method to maintain good tracking performance, using different sensor types and even when starting with unfavourable configurations. We benchmark the statistical performance of the proposed tracking method using a CRB based on a zero-mean random acceleration model. This is a useful but slightly conservative bound.
The code implementation of the method is available here \texttt{TO APPEAR}.

The controller has one main user parameter, $k$. The choice of $K$ depends on the sampling period of the controller relative to the target dynamics, and its maximum value is limited by computational constraints. For all examples, we set $k=7$, which ensures the feasibility of the problem.

\subsection{Measurement models}
\label{sec:measurementmodels}

We consider two types of wireless sensors in the experiments. 
First, we use ranging based on round-trip time measurements, as in \eqref{eq:ranging}, with distance-dependent measurement noise variance for sensor $m$:
\begin{equation}
\sigma^2( \substarget^n, \subssensor^m ) = \frac{\sigma^2_{\text{range}}}{c^2}
\big(1+\lambda \mu^2(\substarget^n, \subssensor^m)\big)
\end{equation}
where $c = 3\cdot 10^{8}$ [m/s], $\sigma_{\text{range}}=1$ [m] and  $\lambda = 0.01$ are constants.
Secondly, we use sensors that measure Doppler frequency shifts \cite{doppler} as described in \eqref{eq:doppler} with a carrier frequency $f_c$   of $2.3 \text{ GHz}$. Here we consider a  noise model for sensor $m$:
\begin{equation}
    \sigma^2(\substarget^n , \subssensor^m) \equiv \sigma^2_{\text{Doppler}}
\end{equation} with a standard deviation of $1~\text{[Hz]}$. The sensors are assumed to be calibrated such that $\sigma_{\text{range}}$ and $\sigma_{\text{Doppler}}$ are known.


\subsection{Three ranging sensors tracking in 2D space}
\begin{figure*}[!htb]
    \centering
    \subfloat[]{\includegraphics[width=0.45\linewidth]{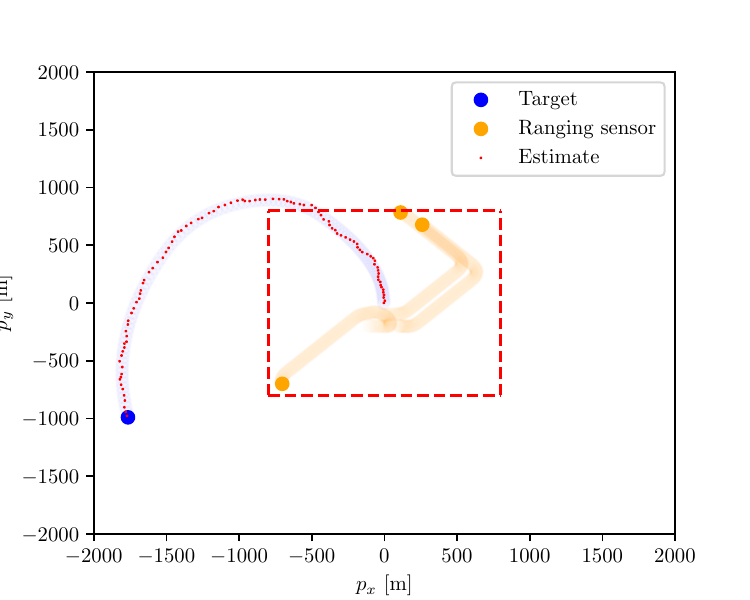}\label{fig:one-target-ranging-trajectory}}
    \subfloat[]{\includegraphics[width=0.45\linewidth]{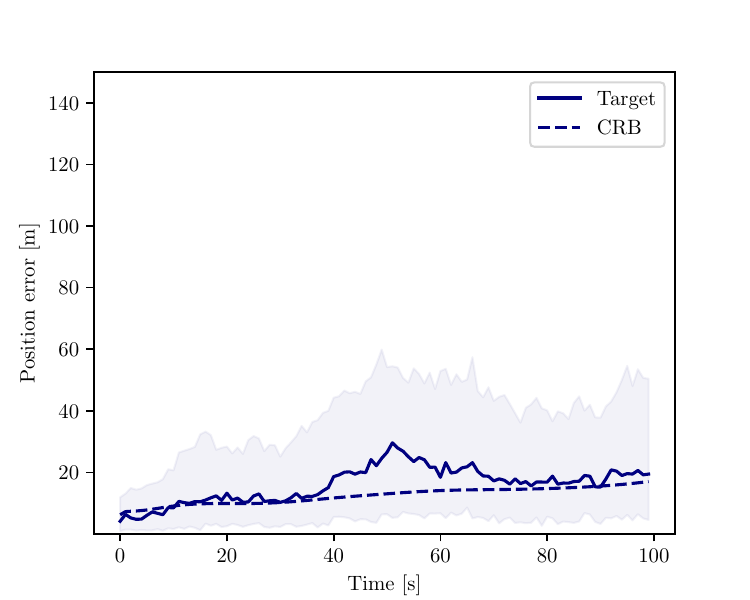}\label{fig:one-target-ranging-err}}
    \caption{
    (a) Ranging sensors tracking a single target in 2D space. (b) The resulting tracking position errors over time. $\boundctr = 2 \text{ m/$\text{s}^2$}$ and $\boundadv= 5 \text{ m/$\text{s}^2$}$.}
    \end{figure*}
We begin with a simple scenario in two-dimensional space ($d=2$) using only ranging measurements, for which we have some intuition. We know that a target position is identifiable provided $M=3$ sensors are well separated and that the achievable accuracy improves by maintaining a large angle of separation between them. At the same time, an increased distance to the target affects the signal-to-noise ratio adversely. Therefore the tracking system must find an appropriate balance between these factors.

To better illustrate the controller's performance, we initialize the sensors at the same location for two different scenarios. In the first scenario, depicted in Figure~\ref{fig:one-target-ranging-trajectory}, the target moves from the center of the perimeter to the bottom left outside. The sensors quickly respond by adjusting their positions. Once sufficiently separated, one sensor moves in the direction of the target, while the other two move orthogonally to the first sensor so as to improve the target resolvability. This dynamic sensor configuration consistently maintains relatively low tracking errors.

The second scenario, shown in Figure~\ref{fig:two-target-ranging-sensor}, is more challenging, as an additional target enters the perimeter from the outside. Initially, the sensors are poorly positioned relative to the new target, resulting in a highly biased estimate. However, the tracking system adapts dynamically, and as the sensor configuration improves, it successfully corrects these systematic errors.

\begin{figure*}[!htb]
    \centering
    \subfloat[]{\includegraphics[width=0.45\linewidth]{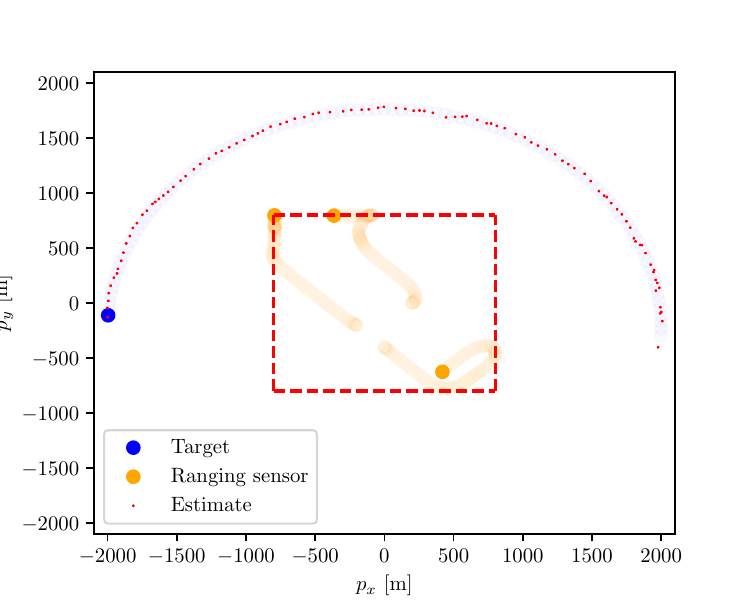}\label{fig:half-circle-trajectory}}
    \subfloat[]{\includegraphics[width=0.45\linewidth]{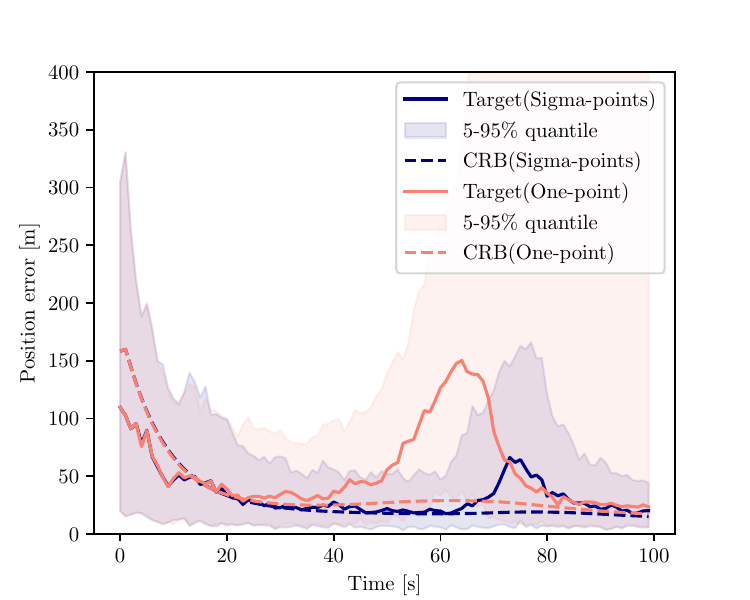}\label{fig:half-circle-err}}
    \caption{
    (a) Ranging sensors tracking a single target in 2D space.  (b) Comparison between robust sigma-point based method \eqref{eq:controlproblem_relax} and nonrobust point-prediction based method \eqref{eq:controlproblem_simple} in the scenario of Fig.~\ref{fig:half-circle-trajectory}, using 100 Monte Carlo simulations. The solid curve denotes the median values. $\boundctr = 2 \text{ m/$\text{s}^2$}$ and $\boundadv= 5 \text{ m/$\text{s}^2$}$.}
    \end{figure*}

We now study the robustness formulation used in \eqref{eq:controlproblem_relax}. What is the gain of averaging over multiple possible target trajectories as compared to a simpler method that uses a point prediction of each trajectory. That is,
\begin{equation}
    \seqctr^{**}_\step =
 \argmin_{\seqctr_\step \in \setctr_{\infty} \cap \setctr_c} 
 \sum^N_{n=1} g(\estimator^n_{t}, \mathbf{0}, \dots, \mathbf{0}, \seqctr_\step )
\label{eq:controlproblem_simple}
\end{equation}
where $\widehat{\adversarial}^n_{i} \equiv\zeromat$ is the optimal point prediction of the future acceleration vectors. We consider the challenging scenario depicted in Figure~\ref{fig:half-circle-trajectory} and compare the tracking errors of the robust and certainty-equivalent systems over 100 Monte Carlo runs. As shown in Figure~\ref{fig:half-circle-err}, while the expected CRBs are comparable, the robust formulation \eqref{eq:controlproblem_relax} significantly reduces the tracking error tails, particularly during the latter part of the tracking phase (approximately $t \geq 50$).

\subsection{Six Doppler sensors tracking a single target}

We now increase the complexity of the tracking problem by considering a target moving in three-dimensional space ($d=3$) and using only Doppler frequency shift measurements. Using the states \eqref{eq:states_def} in  \eqref{eq:doppler}, we see that each Doppler sensor effectively measures the angle between the target direction vector, $(\pos^n - \widetilde{\pos}^m)/\| \pos^n - \widetilde{\pos}^m \|$, and the velocity difference, $(\velo^n - \widetilde{\velo}^m)$. It is, therefore, a nontrivial task to dynamically control the sensors to maintain a good tracking performance for a moving target.

\begin{figure*}
    \centering

     \subfloat[]{\includegraphics[width=0.45\linewidth]{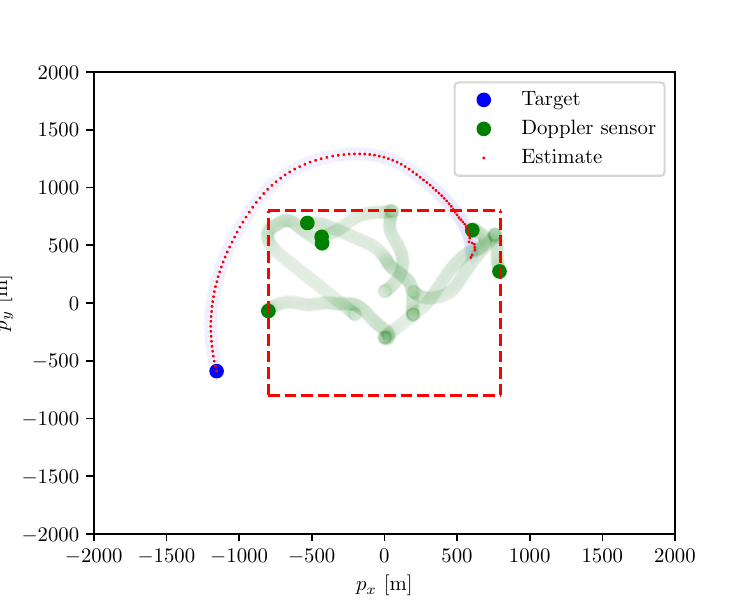}\label{fig:one-target-doppler-trajectory-1}}
     \subfloat[]{\includegraphics[width=0.45\linewidth]{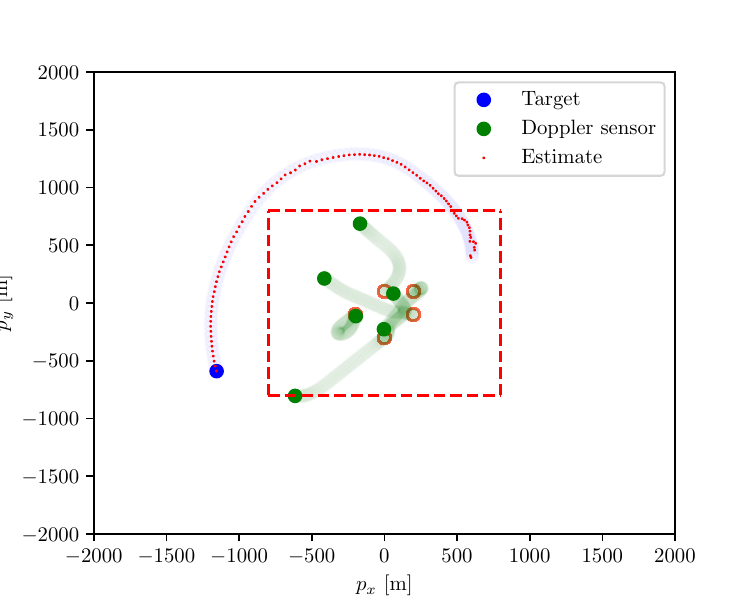}\label{fig:one-target-doppler-trajectory-2}}
    
    \caption{Tracking using $M=6$ mobile Doppler sensors. $\boundctr = 2 \text{ m/$\text{s}^2$}$ and $\boundadv= 5 \text{ m/$\text{s}^2$}$. (a) Mobile sensor control is on at all times. (b) Control is turned off during $t \leq 50$, the red circles indicate locations when off.}
    \label{fig:one-target-doppler}
\end{figure*}

\begin{figure*}[!t]
    \centering
    \subfloat[]{\includegraphics[width=0.45\linewidth]{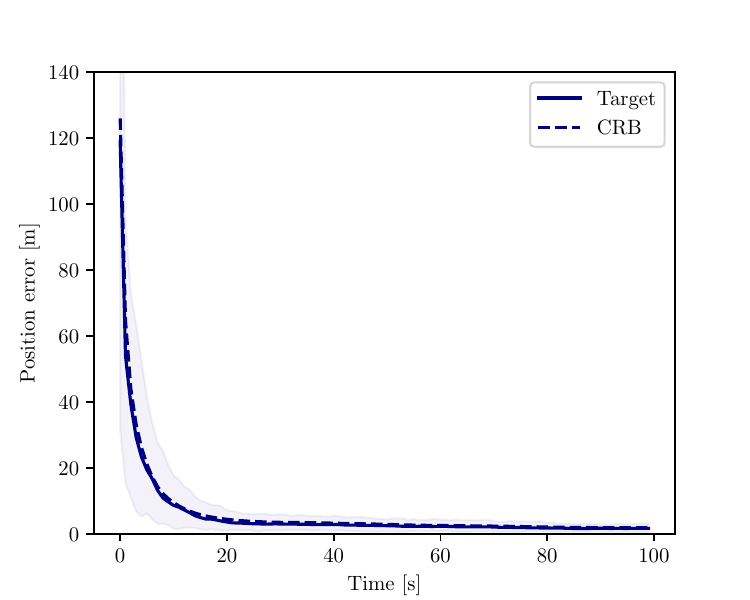}\label{fig:one-target-doppler-perr}}
    \subfloat[]{\includegraphics[width=0.45\linewidth]{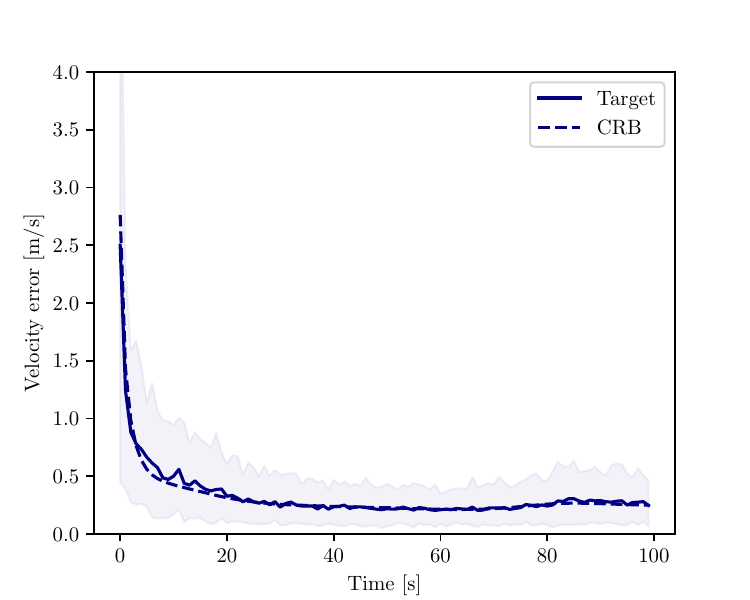}\label{fig:one-target-doppler-verr}}
    
    \subfloat[]{\includegraphics[width=0.45\linewidth]{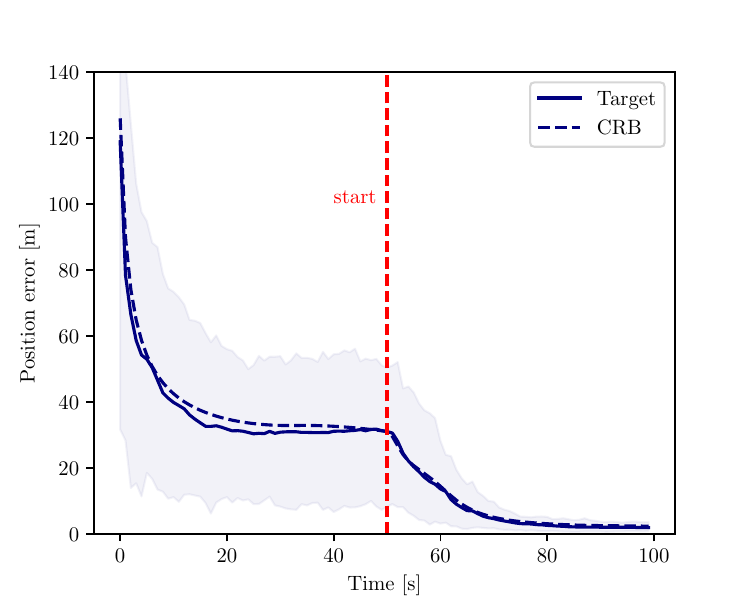}\label{fig:one-target-doppler-perr-2}}
    \subfloat[]{\includegraphics[width=0.45\linewidth]{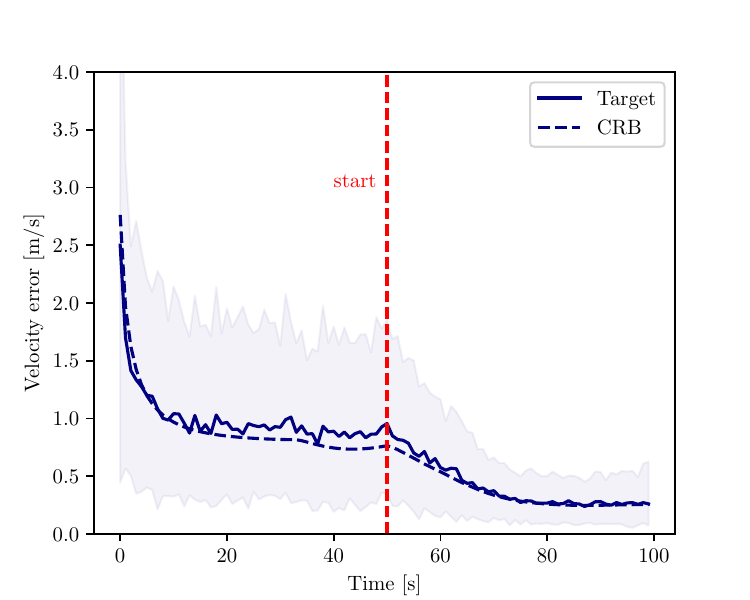}\label{fig:one-target-doppler-verr-2}}
    \caption{(a) and (b) show the position and velocity errors when the control algorithm starts at $\step=0$, while (c) and (d) illustrates the case where mobile sensor control begins at $\step=50$.}
    \label{fig:one-target-doppler-err}
\end{figure*}

We consider $M=6$ Doppler sensors as shown in Figure~\ref{fig:one-target-doppler}.  The target starts moving from the top-right corner of the perimeter, following an arc-shaped trajectory along the outer boundary toward the bottom-left. Along this path, the angle between the target's direction vector and the velocity difference changes rapidly, making the sensor responses more complex. The resulting tracking errors are shown in Figure~\ref{fig:one-target-doppler-err}. For comparison, in a second instance (Figure~\ref{fig:one-target-doppler-trajectory-2}), we disable sensor mobility during $\step \leq 50$. During this period, the CRB and tracking error stabilizes at a plateau but remain noticeably higher than when sensor motion starts at $\step = 0$. In both cases, however, the CRB and tracking error eventually converge to similar levels by  $\step = 100$.


\subsection{Four mixed sensors tracking a single target}
\begin{figure*}[!htb]
    \centering
    \subfloat[]{\includegraphics[width=0.45\linewidth]{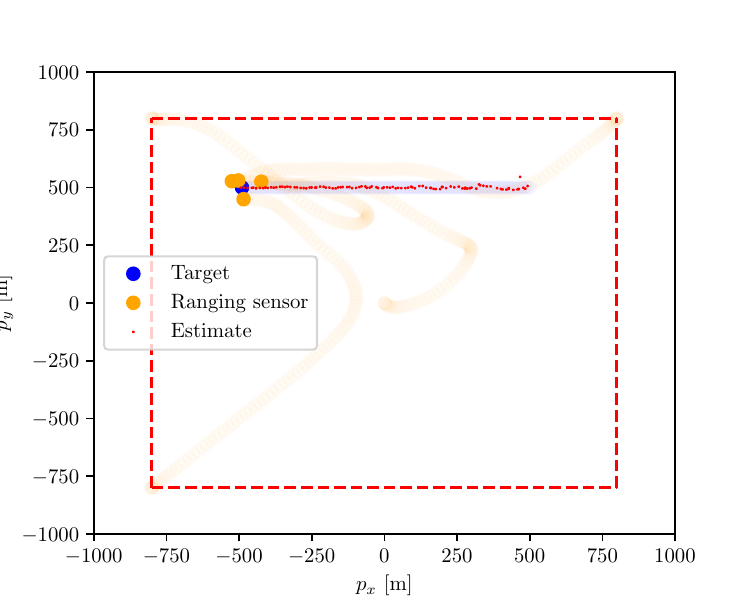}\label{fig:one-target-ranging-3D}}
    \subfloat[]{\includegraphics[width=0.45\linewidth]{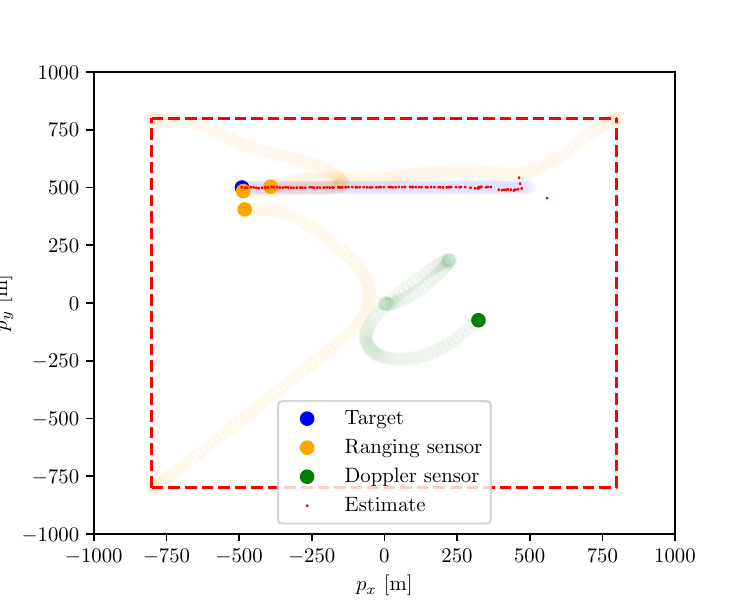}\label{fig:one-target-mix}}
    \caption{(a) Tracking a single target using four ranging sensors. (b) Tracking the same target as Figure~\ref{fig:one-target-ranging-3D} using three ranging sensors and one Doppler sensor. $\boundctr = 2 \text{ m/$\text{s}^2$}$ and $\boundadv= 5 \text{ m/$\text{s}^2$}$.}
    \label{fig:one-target-mix-vs-range}
\end{figure*}

 \begin{figure*}[!htb]
    \centering
    \subfloat[]{\includegraphics[width=0.45\linewidth]{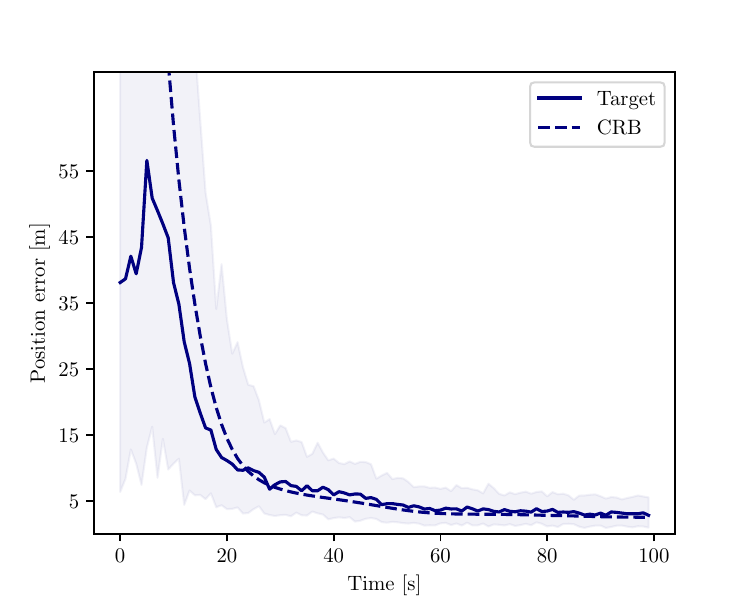}\label{fig:one-target-ranging-3D-perr}}
    \subfloat[]{\includegraphics[width=0.45\linewidth]{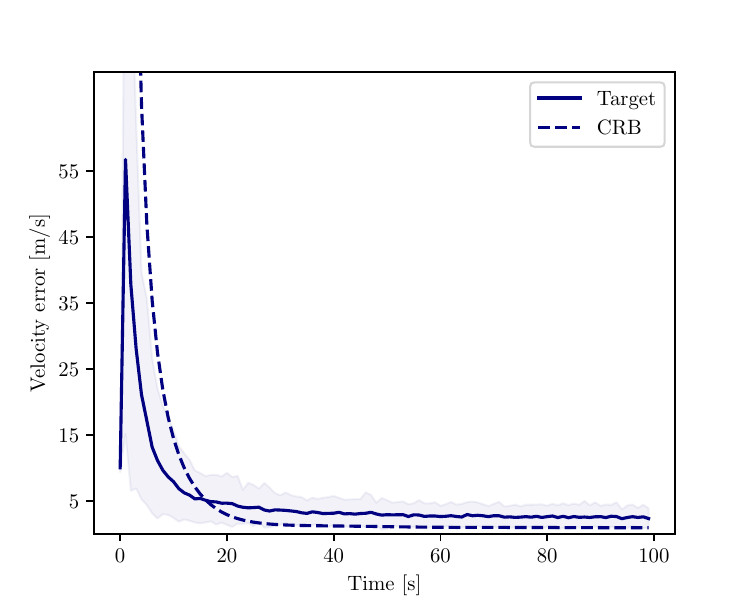}\label{fig:one-target-ranging-3D-verr}}
    
    \subfloat[]{\includegraphics[width=0.45\linewidth]{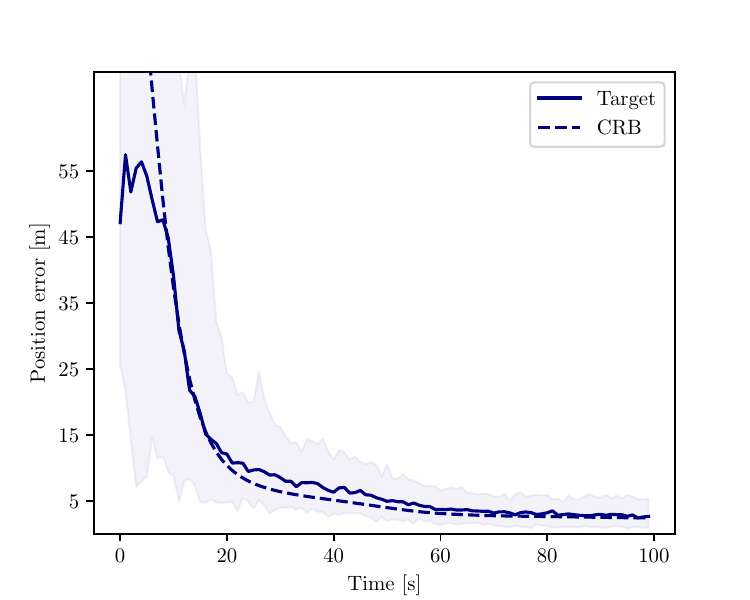}\label{fig:one-target-mix-perr}}
    \subfloat[]{\includegraphics[width=0.45\linewidth]{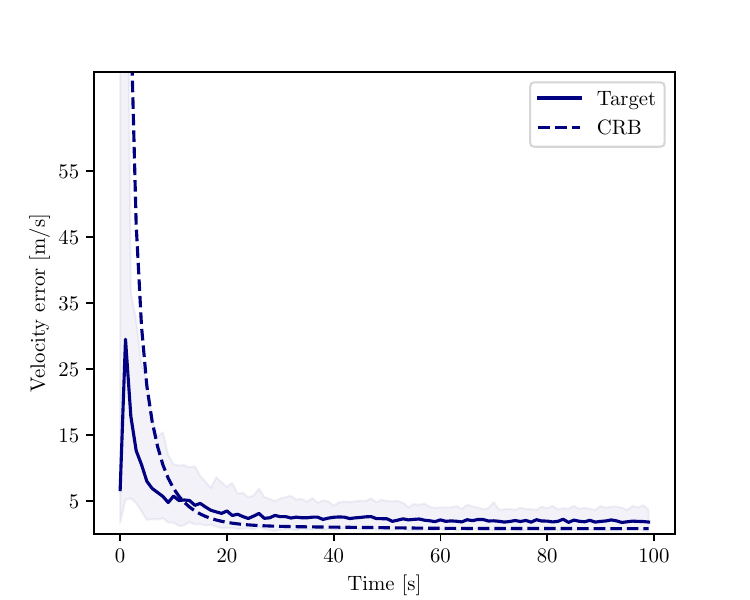}\label{fig:one-target-mix-verr}}
    \caption{ (a) and (b), position and velocity errors when tracking with four ranging sensors as shown in Figure \ref{fig:one-target-ranging-3D}. (c) and (d), position and velocity errors when tracking with three ranging sensors and one Dopper sensor as shown in Figure \ref{fig:one-target-mix}, using 100 Monte Carlo simulations.}
    \label{fig:one-target-mix-vs-range-err}
\end{figure*}

Additional complexity is introduced into the scenario above when we use a mix of sensor types. Figure~\ref{fig:one-target-ranging-3D} illustrates a simple case where a target follows a slow, uniform rectilinear motion. Three ranging sensors are initially positioned at different corners of the perimeter, while a fourth is placed at the center of the area. The sensor movements are straightforward to interpret—sensors will approach the target, aligning to achieve a good dilution of precision. Next, we replace one ranging sensor with a Doppler sensor, as shown in Figure~\ref{fig:one-target-mix}. Unlike ranging sensors, Doppler sensors have a distance-independent noise model. As a result, the Doppler sensor follows an intricate path, deviating from the target, take into account both the angle and magnitude of the velocity difference to the target. As shown in Figure~\ref{fig:one-target-mix-vs-range-err}, both position and velocity errors in the two cases quickly stabilize. Notably, in the mixed-sensor setup, the velocity error decreases to a lower level compared to the case with only ranging sensors.

\section{CONCLUSION}

We have considered the problem of tracking multiple targets using mobile sensors of (possibly) different types.
The proposed tracking system employs minimal assumptions about the target dynamics, merely that their accelerations are bounded. The sensor control law uses a maximum likelihood estimator to determine the sensor accelerations so as to maximize target resolvability, while taking into account the least favourable target trajectories. The objective is posed using the expected Cram\'{e}r-Rao bound. The mobile sensor control law is implemented using an efficient interior point method. The resulting method is illustrated in a series of experiments involving both round-trip time based ranging and Doppler frequency shift measurements. 

The tracking system exhibits an intuitive behaviour in simple settings, such as a single target with range-only sensors. However, for mixed sensor types or multiple targets, the resulting sensor trajectories are nontrivial. We showed that the proposed joint estimation and control solution leads to robust tracking performance even when starting from highly challenging initial sensor configurations.

\section*{ACKNOWLEDGMENT}

The authors would like to thank Dr. Torbjörn Wigren for helpful comments and advice on the simulation scenarios.
\bibliographystyle{plainnat}
\bibliography{ref}
\end{document}